# A Relativistic Theory of Consciousness (shortened version)

Nir Lahav[1*] Zachariah A. Neemeh[2,3]

- [1]Department of Physics, Bar-Ilan University, Ramat Gan, Israel
- [2]Department of Philosophy, The University of Memphis, Memphis, TN, United States
- [3]Institute for Intelligent Systems, The University of Memphis, Memphis, TN, United States

**Abstract**

This paper is a shortened version of the full paper that was published in the journal Frontiers of Psychology in May 2022.

In recent decades, the scientific study of consciousness has significantly increased our understanding of this elusive phenomenon. Yet, despite critical development in our understanding of the functional side of consciousness, we still lack a fundamental theory regarding its phenomenal aspect. There is an "explanatory gap" between our scientific knowledge of functional consciousness and its "subjective," phenomenal aspects, referred to as the "hard problem" of consciousness. The phenomenal aspect of consciousness is the first-person answer to "what it's like" question, and it has thus far proved recalcitrant to direct scientific investigation. The question of how the brain, or any cognitive system, can create conscious experience out of neural representations poses a great conundrum to science. Naturalistic dualists argue that it is composed of a primitive, private, nonreductive element of reality that is independent from the functional and physical aspects of consciousness. Illusionists, on the other hand, argue that it is merely a cognitive illusion, and that all that exists are ultimately physical, nonphenomenal properties. We contend that both the dualist and illusionist positions are flawed because they tacitly assume consciousness to be an absolute property that doesn't depend on the observer. We developed a conceptual and a mathematical argument for a relativistic theory of consciousness in which a system either has or doesn't have phenomenal consciousness *with respect to some observer*. According to the theory, Phenomenal consciousness is neither private nor delusional, just relativistic. In the frame of reference of the cognitive system, it will be observable (first-person perspective) and in other frame of reference it will not (third-person perspective). These two cognitive frames of reference are both correct, just as in the case of an observer that claims to be at rest while another will claim that the observer has constant velocity. Given that consciousness is a relativistic phenomenon, neither observer position can be privileged, as they both describe the same underlying reality. Based on relativistic phenomena in physics we developed a mathematical formalization for consciousness which bridges the explanatory gap and dissolves the hard problem. Given that the first-person cognitive frame of reference also offers legitimate observations on consciousness, we conclude by arguing that philosophers can usefully contribute to the science of consciousness by collaborating with neuroscientists to explore the neural basis of phenomenal structures.

As one of the most complex structures we know of nature, the brain poses a great challenge to us in understanding how higher functions like perception, cognition, and the self arise from it. One of its most baffling abilities is its capacity for conscious experience (van Gulick, 2014). Thomas Nagel (1974) suggests a now widely-accepted definition of consciousness: a being is conscious just if there is "something that it is like" to be that creature, i.e., some subjective way the world seems or appears from the creature's point of view. For example, if bats are conscious, that means there is something it is like for a bat to experience its world through its echolocational senses. On the other hand, under deep sleep (with no dreams) humans are unconscious because there is nothing it is like for humans to experience their world in that state.

In the last several decades, consciousness has transformed from an elusive metaphysical problem into an empirical research topic. Nevertheless, it remains a puzzling and thorny issue for science. At the heart of the problem lies the question of the brute phenomena that we experience from a first-person perspective—e.g., what it is like to feel redness, happiness, or a thought. These qualitative states, or qualia, compose much of the phenomenal side of consciousness. These qualia are arranged into spatial and temporal patterns and formal structures in phenomenal experience, called eidetic or transcendental structures. For example, while qualia pick out how a specific note sounds, eidetic structures refer to the temporal form of the whole melody. Hence, our inventory of the elusive properties of phenomenal consciousness includes both qualia and eidetic structures.

One of the central aspects of phenomenal features is privacy. It seems that my first-person feeling of happiness, for example, cannot be measured from any other third-person perspective. One can take indirect measurements of my heart rate or even measure the activity in the brain networks that create the representation of happiness in my mind, but these are just markers of the feeling that I have, and not the feeling itself (Block, 1995). From my first-person perspective, I don't feel the representation of happiness. I just feel happiness. Philosophers refer to this feature of phenomenal consciousness as 'transparency': we seem to directly perceive things, rather than mental representations, even though mental representations mediate experience. But this feeling cannot be directly measured from the third-person perspective, and so is excluded from scientific inquiry. Yet, let us say that we identify a certain mental representation in a subject's brain as 'happiness.' What justifies us in calling it 'happiness' as opposed to 'sadness?' Perhaps it is correlated with certain physiological and behavioral measures. But, ultimately, the buck stops somewhere: at a subject's phenomenological report otherwise, we cannot know what this representation represents (Gallagher & Zahavi, 2020). If we categorize the representation as 'happiness' but the subject insists they are sad, we are probably mistaken, not the subject. Phenomenal properties are seemingly private and, by some accounts, even beyond any physical explanation. For being physical means being public and measurable and is epistemically inconsistent with privacy. Take an electron, for example. We know it is physical because we can measure it with e.g., a cathode ray tube. If electrons were something that only I could perceive (i.e., if they were private), then we would not include them as physical parts of the scientific worldview. Affective scientists, for example, measure aspects of feeling all the time, like valence and arousal, but measuring these is not the same as measuring the feeling of happiness itself. What affective scientists measure are outcomes of the feeling, while the feeling itself is private and not measurable by the scientists. To date, it is not clear how to bridge the "explanatory gap" (Levine, 1983) between "private" phenomenal features and public, measurable features (e.g., neurocomputational structures), leaving us stuck in the "hard problem" of consciousness (Chalmers, 1996). This gap casts a shadow on the possibility for neuroscience to solve the hard problem because all explanations will always remain within a third-person perspective (e.g., neuronal firing patterns and representations), leaving the first-person perspective out of the reach of neuroscience. This situation divides consciousness into two separate aspects, the functional aspect and the phenomenal one (Block, 1995). The functional aspect ('functional consciousness'), is the objectively observable aspect of consciousness (Franklin et al., 2016; Kanai et al., 2019). (In that sense, it's similar to Ned Block's definition of access conscious, but with less constraints. All phenomenal consciousness has a functional aspect and vice versa, whereas for Block this strict equivalence doesn't hold [Block, 2011]). But the subjective aspect (phenomenal consciousness) is not directly observable except on the part of the person experiencing that conscious state. As we saw above, because they are private, phenomenal properties are distinct from any cognitive and functional property (which can be publicly observable). Any theory of consciousness should explain how to bridge this gap—How can functional, public, aspects give rise to phenomenal, private aspects?

Nevertheless, in recent decades, consciousness has become increasingly amenable to empirical investigation by focusing on its functional aspect, finally enabling us to begin to understand this enigmatic phenomenon. For example, we now have good evidence that consciousness doesn't occur in a single brain area. Rather, it seems to be a global phenomenon in widespread areas of the brain (Baars, 1988; Mashour et al., 2020; Varela et al., 2001).

In studying *functional consciousness*, we take consciousness to be a form of information processing and manipulation of representations, and we trace its functional or causal role within the cognitive system. Widely successful theories, such as global workspace theory (Baars, 1988; Dehaene, 2014; Franklin et al., 2016), attention schema theory (Graziano, 2019), recurrent processing theory (Fahrenfort et al., 2007; Lamme, 2006), and integrated information theory (Oizumi et al., 2014; Tononi, 2008) are virtually premised on this information processing account. Despite our advances in the study of functional consciousness, we still lack a convincing way to bridge the explanatory gap to phenomenal consciousness. The questions, "Why does it feel like anything at all to process information?", and "How can this feeling be private?" still remain controversial (for a criticism about Integrated Information Theory, see full paper, introduction section). The hard problem primarily remains a philosophical rather than scientific question. Phenomenal consciousness must be the ultimate reference point for any scientific theory of consciousness. Ultimately, theories of consciousness as information processing, i.e., theories of functional consciousness, only approximate full-blown consciousness by abstracting away its phenomenal features. But they must ultimately refer to phenomenal features in order to give a full explanation of consciousness. Otherwise, there are no grounds for labeling them theories of consciousness, rather than theories of cognition or global informational access. As David Chalmers (1995) has put it, "the structure and dynamics of physical processes yield only more structure and dynamics, so structures and functions are all we can expect these processes to explain." The structure and dynamics of information tell us nothing of the story of how one gets from these public structures and dynamics to our private phenomenal experience (and how there can be an equality between these opposite properties). This is known as the 'structure and dynamics argument' (Alter, 2016; D. J. Chalmers, 2003; Mindt, 2017), stating that structure and dynamics alone are not enough to account for consciousness. This raises a concern about whether any physicalist theory can solve the hard problem of consciousness. In the next sections, however, we will show that physicalism is broader than describing only structures and dynamics, and that we can use this fact in order to solve the hard problem.

Views about the relation between phenomenal and functional consciousness exist across a spectrum. On one end, illusionists seek to erase the hard problem by referring only to functional consciousness, taking phenomenal properties to be cognitive illusions. On the other end, naturalistic dualists or panpsychists seek to promote phenomenal consciousness as a fundamental, nonmaterial, and undecomposable constituent of nature (Chalmers, 2017). The controversy over phenomenal consciousness can be traced to one central problem: naturalization. The project of naturalization involves taking folk psychological concepts and subjecting them to physical laws and empirical scrutiny (Hutto, 2007). Illusionists take the current scientific approach to consciousness and argue that this eliminates the messy problem with supposedly private, immaterial qualia. According to them, functional consciousness generates an illusion of special phenomenal properties, which create the persistent "user illusion" (Dennett, 1991) of a first-person perspective. Dualists start from the same problem of naturalization, but take it that phenomenal consciousness is simply not amenable to third-person scientific inquiry due to its sui generis properties. What is needed, according to the naturalistic dualist, is an expanded understanding of what counts as 'natural.' "Given that reductive explanation fails, *nonreductive* explanation is the natural choice" (Chalmers, 2017, p. 359). Chalmers proposes that consciousness is a fundamental property, something like the strong nuclear force that is irreducible to other forces. A complete ontology of the natural world simply must include phenomenal consciousness as a basic, undecomposable constituent.

**The zombie argument and the paradox of phenomenal judgment**

Chalmers (1996) discusses the logical possibility of a *zombie*, a being physically, cognitively, and behaviorally identical to a human, but lacking phenomenal consciousness altogether. One would think that such a creature would be dull, like a robot with basic automated responses, but this is not the case. Because a zombie is physically identical to a human, it means that it has the same cognitive system as us: a system that gets inputs from the environment, processes them, and creates behavior responses. In fact, there are no differences between the human and the zombie's cognitive dynamics, representations, and responses. But, for the zombie, there is nothing that is like to do all these processes. According to Chalmers (1996), the zombie has *phenomenal judgments*. This concept is very important for our argument, so let's examine it a bit. Phenomenal judgements are higher-order cognitive functions that humans and zombies have in common. Humans are aware of their experience and its contents and they can form judgments about it (e.g., when we think 'There is something red'), then, usually, they are led to make claims about it. These various judgments in the vicinity of consciousness are phenomenal judgments. They are not phenomenal states themselves, but they are about phenomenology. Phenomenal judgments are often reflected in claims and reports about consciousness, but they start as a mental process. Phenomenal judgments are themselves cognitive acts that can be explained by functional aspects like the manipulation of mental representations. That's why zombies also have phenomenal judgements. We can think of a judgment as what is left of a belief after any associated phenomenal property is subtracted. As a result, phenomenal judgements are part of the functional aspect of consciousness. As Chalmers puts it (1996, p.174):

"Judgments can be understood as what I and my zombie twin have in common. A zombie does not have any conscious experience, but he claims that he does. My zombie twin judges that he has conscious experience, and his judgments in this vicinity correspond one-to-one to mine. He will have the same form, and he will function in the same way in directing behavior as mine… Alongside every conscious experience there is a content-bearing cognitive state. It is roughly information that is accessible to the cognitive system, available for verbal report, and so on."

In other words, phenomenal judgments can be described as the representations of a cognitive system that bear content about phenomenology (which are not necessarily linguistic representations. E.g., such representation can be a representation of the color of the apple). In this paper, we will identify functional consciousness with the creation of phenomenal judgments.

As a result of the zombie's capacity to create phenomenal judgements, we reach a peculiar situation: The zombie has functional consciousness, i.e., all the physical and functional conscious processes studied by scientists, such as global informational access. But there would be nothing it is like to have that global informational access and to be that zombie. All that the zombie cognitive system requires is the capacity to produce phenomenal judgments that it can later report. For example, if you asked it if it sees a red rose in front of it, using information processing, it might respond, "Yes, I'm definitely conscious of seeing a red rose", although it is ultimately mistaken and there is truly nothing that is like for the zombie to see that rose. In order to produce this phenomenal judgment, despite having no phenomenal consciousness, the zombie cognitive system needs representations and a central system with direct access to important information enabling it to generate behavioral responses. It needs direct access to perceptual information, a concept of self to differentiate itself from the world, an ability to access its own cognitive contents, and the capacity to reflect. Such a cognitive system could presumably reason about its own perceptions. It would report that it sees the red rose, and that it has some property over and above its structural and functional properties—phenomenal consciousness. Of course, this report would be mistaken. It is a paradoxical situation in which functional consciousness creates phenomenal judgments without the intervention of phenomenal consciousness—yet phenomenal judgments are purportedly about phenomenal consciousness. This paradox of phenomenal judgment (Chalmers, 1996) arises because of the independence of phenomenal consciousness from physical processes. The hidden assumption here is that consciousness is private. Consequently, it is not possible to measure it. It seems that one aspect of consciousness (physical, functional consciousness) can come without the other (phenomenal consciousness).

In order to solve this paradox, we need to explain two aspects of consciousness: How there could be natural phenomena that are private and thus independent of physical processes (or how come they *seem* private), and what the exact relationship between cognitive content and phenomenal consciousness is.

**The relativistic approach**

A common thread connecting both extremes of dualism and illusionism is that both assume that phenomenal consciousness is an absolute phenomenon, wherein an object *O* evinces either property *P* or ¬*P*. We will show that we need to abandon this assumption. The relativistic principle in modern physics posits a universe in which for many properties an object *O* evinces either property *P* or ¬*P with respect to some observer X*. In such a situation, there is no one answer to the question of whether object *O* has property *P* or not. In Lahav and Neemeh (2022) we proposed a novel relativistic theory of consciousness in which consciousness is not an absolute property but a relative one. In this shortened version of the full paper we introduce the conceptual aspects of the theory and their outcomes. This approach eschews both extremes of illusionism and dualism. The relativistic theory of consciousness will show that phenomenal consciousness is neither an illusion nor a unique fundamental property of the universe. It will give a coherent answer to the question of the (supposed) privacy of phenomenal consciousness, will bridge the explanatory gap, and will provide a solution to the hard problem based in relativistic physics. General notions of this approach can be found in some dual aspect monisms such as Max Velmans' reflexive monism. Here, however, we develop a physical theory of consciousness as a relativistic phenomenon and formalize the perspectival relations in light of the relativistic principle. To do that, in the following section, we will develop more formally the *relativistic principle* and introduce the *equivalence principle of consciousness*.

In physics, relativity means that different observers from different frames of reference will nevertheless measure the same laws of nature. If, for example, one observer is in a closed room in a building and the other observer is in a closed room in a ship (one moving smoothly enough on calm water), then the observer in the ship would not be able to tell whether the ship is moving or stationary. Each will obtain the same results for any experiment that tries to determine whether they are moving or not. For both of them, the laws of nature will be the same, and each will conclude that they are stationary. For example, if they throw a ball towards the room's ceiling, each will determine that the ball will return directly into their hands (because the ship moves with constant velocity and because of Newton's First Law, the ball will preserve the velocity of the ship while in the air, and will propagate forward with the same pace as the ship. As a result, it will fall directly into the observer's hands). There will be no difference in the results of each observer's measurements, trying to establish whether they are stationary or not. They will conclude that they have the same laws of nature currently in force, causing the same results. These results will be the results of a stationary observer and thus both of them will conclude that they are at rest. Because each of them will conclude that they are the stationary one, they will not agree about one another's status. Each of them will conclude that the other is the one that moves. (Common sense will tell the observer in the ship that they are the one moving, but imagine an observer locked on the ship in a room with no windows. Such an observer cannot observe the outside world. This kind of observers will conclude that they are stationary because velocity is relativistic.)

To state that consciousness is a relativistic phenomenon is to state that there are observers in different *cognitive frames of reference,* yet they will nevertheless measure the same laws of nature currently in force and the same phenomenon of consciousness in their different frames. We will start with an equivalence principle between a conscious agent, like a human being, and a zombie agent, like an advanced artificial cognitive system. As a result of this equivalence, we will show that if the relativistic principle is true, then zombies are not possible. Instead, every purported zombie will actually have phenomenal consciousness and any system with adequate functional consciousness will exhibit phenomenal consciousness from the first-person cognitive frame of reference. Others have similarly claimed that zombies are physically impossible (Brown, 2010; Dennett, 1995; Frankish, 2007; Nagel, 2012), but our aim here is to show why that is according to the relativistic principle. As a result of this equivalence, observations of consciousness fundamentally depend on the observer's cognitive frame of reference. The first-person cognitive frame of reference is the perspective of the cognitive system itself (Solms, 2021). The third-person cognitive frame of reference is the perspective of any external observer of that cognitive system. Phenomenal consciousness is only seemingly private because in order to measure it one needs to be in the appropriate cognitive frame of reference. It is not a simple transformation to change from a third-person cognitive frame of reference to the first-person frame, but in principle it can be done, and hence phenomenal consciousness isn't private anymore.

We will show that from its own first-person cognitive frame of reference, the observer will observe phenomenal consciousness, but any other observer in a third-person cognitive frame of reference will observe only the physical substrates that underlie qualia and eidetic structures. The illusionist mistake is to argue that the third-person cognitive frame of reference is *the* proper perspective. To be clear, the first-person cognitive frame of reference is still a physical location in space and time (it is not immaterial); it is just the position and the dynamics of the cognitive system itself. As we will see, this is the position from which phenomenal consciousness can be observed. The principle of relativity tells us that there is no privileged perspective in the universe. Rather, we will get different measurements depending on the observer's position. Since consciousness is relativistic, we get different measurements depending on whether the observer occupies or is external to the cognitive system in question. Both the first-person and third-person cognitive frames of reference describe the same reality from two different points of view, and we cannot prefer one point of view upon the other (Solms, 2021). The dualist mistake is to argue that phenomenal consciousness is private. For any relativistic phenomenon there is a formal transformation between the observers of different frames of reference, meaning that both frames can be accessible to every observer with the right transformations. Consciousness as a relativistic phenomenon also has such transformation rules. We will describe the transformations between first-person (i.e., phenomenological) and third-person (i.e., laboratory point of view) cognitive frames of reference. There are several consequences of these transformations. First, qualia and eidetic structures are not private. Rather, they only *appear* private, because in order to measure them one needs to be in the appropriate cognitive frame of reference, i.e., within the perspective of the cognitive system in question. We can use these transformations to answer questions like, "What is it like to be someone else?". Because of the transformations, results that we obtain from third-person methodology should be isomorphic to first-person structures. Isomorphism between two elements means that they have the same mathematical form and there is a transformation between them that preserves this form. Equality is when two objects are exactly the same, and everything that is true about one object is also true about the other. However, an isomorphism implies that everything that is true about *some* properties of one object's structure is also true about the other. We will show that this is the case with measurements obtained from first-person and third-person frames of reference. We will show that this isomorphism is a direct result of the relativistic principle and the notion that phenomenal judgements and phenomenal structures are two sides of the same underlying reality. All that separates them are different kinds of measurements (causing different kinds of properties). An unintuitive consequence of the relativistic theory is that the opposite is also true, and first-person structures also bear formal equivalence to third-person structures. We advocate for interdisciplinary work between philosophers and cognitive neuroscientists in exploring this consequence.

**The Principle of Relativity and the Equivalence Principle**

Our task is to establish the equivalence principle of consciousness, namely, that qualitative and quantitative aspects of consciousness are formally equivalent. We start by establishing the equivalence between conscious humans and zombies, and then we expand that equivalence to all structures of functional consciousness. In this shorten version of the paper we will present only the conceptual defense of the equivalence principle (for the mathematical formalization see [the full paper](#)).
We must first present the principle of relativity and the equivalence principle in physics. Later, we will use these examples to develop a new equivalence principle and a new transformation for consciousness. To be more formal than earlier, the principle of relativity is the requirement that the equations describing the laws of physics have the same form in all admissible frames of reference (Møller, 1952). In physics, several relativistic phenomena are well-known, such as velocity and time, and the equivalence principle between uniformly accelerated system and a system under a uniform gravitational field. Let's examine two examples with the help of two observers, Alice and Bob.

In the first example of a simple Galilean transformation, Alice is standing on a train platform and measures the velocity of Bob, who is standing inside a moving train. Meanwhile, Bob simultaneously measures his own velocity. As the train moves with constant velocity, we know that the laws of nature are the same for both Alice's and Bob's frames of reference (and thus that the equations describing the laws of physics have the same form in both frames of reference; Einstein et al., 1923/1952, p. 111). According to his measurements, Bob will conclude that he's stationary, and that Alice and her platform are moving. However, Alice will respond that Bob is mistaken, and that *she* is stationary while Bob and the train are moving. At some point, Alice might say to Bob that he has an illusion that he is stationary and that she is moving. After all, it is not commonsensical that she, along with the platform and the whole world, are moving. Still, although it doesn't seem to make common sense, in terms of physics all of Bob's measurements will be consistent with him being stationary and Alice being the one who moves. In a relativistic universe, we cannot determine who is moving and who is stationary because all experiment results are the same whether the system is moving with constant velocity or at rest. Befuddled, Bob might create an elaborate argument to the effect that his measurement of being stationary is a private measurement that Alice just can't observe. But, of course, both of their measurements are correct. The answer depends on the frame of reference of the observer. Yet, both of them draw mistaken conclusions from their correct measurements. Bob has no illusions, and his measurements are not private. Velocity is simply a relativistic phenomenon and there is no answer to the question of what the velocity of any given body is *without reference to some observer*. Their mistakes are derived from the incorrect assumption that velocity is an absolute phenomenon. As counterintuitive as it may seem, the relativity principle tells us that Bob is not the one who is "really" moving. Alice's perspective is not some absolute, "correct" perspective that sets the standard for measurement, although we may think that way in our commonsense folk physics (Forshaw & Smith, 2014). Alice is moving relative to Bob's perspective, and Bob is moving relative to Alice's perspective. Furthermore, the fact that Alice and Bob agree about all the results from their own measurements means that these two frames of reference are physically equivalent (i.e., they have the same laws of nature in force and cannot be distinguished by any experiment). Because of this physical equivalence, they will not agree about who is at rest and who is moving.

Later, Einstein extended the relativity principle by creating special relativity theory. The Galilean transformation showed that there is a transformation between all frames of reference that have constant velocity relative to each other (inertial frames). Einstein extended this transformation and created the Lorentz transformation, which describes more accurately the relativistic principle. enabling us to move from measurements in one inertial frame of reference to measurements in another inertial frame of reference, even if their velocity is near the speed of light (Einstein, 1905; Forshaw & Smith, 2014). According to the transformation, each observer can change frames of reference to any other inertial frame by changing the velocity of the system. The transformation equation ensures that in each frame we'll get the correct values that the system will measure. For example, the observer will always measure that its own system is at rest and that it is the origin of the axes. Indeed, this is the outcome of the transformation equation for the observer's own system.

The second example comes from Albert Einstein's (1907) observation of the equivalence principle between a uniformly accelerated system (like a rocket) and a system under a uniform gravitational field (like the Earth). Here, Einstein extended the relativity principle even more, not only to constant velocity but also to acceleration and gravity (for local measurements that measure the laws of nature near the observer). He started with two different frames of reference that have the exact same results from all measurements made in their frames of reference. Then he used the relativity principle, concluding that because they cannot be distinguished by any local experiment they are equivalent and have the same laws of nature. Lastly, he concluded that because of the equivalence, we can infer that phenomena happening in one frame of reference will also happen in the other. Now, assume that Alice is skydiving and is in freefall in the Earth's atmosphere, while Bob is floating in outer space inside his spaceship. Although Alice is falling and Bob is stationary in space, both will still obtain the same results of every local experiment they might do. For example, if they were to release a ball from their hand, both of them will measure the ball floating beside of them (in freefall, all bodies fall with the same acceleration and appear to be stationary relative to one another—this is why movies sometimes use airplanes in freefall to simulate outer space). Although they are in different physical scenarios, both will infer that they are floating at rest. Einstein concluded that because they would measure the same results, there is an equivalence between the two systems. In an equivalence state, we cannot distinguish between the systems by any measurement, and thus a system under either gravity or acceleration will have the same laws of nature in force, described by the same equations regardless of which frame it is.

Because of this equivalence, we can infer physical laws from one system to the other. For example, from knowing about the redshift effect of light in accelerated systems with no gravitational force, Einstein predicted that there should be also a redshift effect of light in the presence of a gravitational field as if it were an accelerated system. This phenomenon was later confirmed (Pound & Rebka Jr, 1960).

Next, we argue that phenomenal consciousness is a relativistic physical phenomenon just like velocity. This allows us to dissolve the hard problem by letting consciousness be relativistic instead of absolute. Moreover, we develop a similar transformation for phenomenal properties between first-person and third-person perspectives. The transformation describes phenomenal consciousness with the same form in all admissible cognitive frames of reference and thus satisfying the relativistic principle (see mathematical section in the full paper).

**The Equivalence Principle of Consciousness**

Before we begin the argument, it is essential to elaborate on what 'observer' and 'measurement' mean when applying relativistic physics to cognitive science. In relativity theory, an observer is a frame of reference from which a set of physical objects or events are being measured locally. In our case, let's define a '*cognitive frame of reference*'. This is the perspective of a specific cognitive system from which a set of physical objects and events are being measured. Cognitive frame of reference is being determined by the dynamics of the cognitive system (for more details, see section 2.3 in the full paper).

We use the term 'measurement' in as general a way as possible, from a physical point of view, such that a measurement can occur between two particles like an electron and a proton. Measurement is an interaction that causes a result in the world. The result is the measured property, and this measured property is new information in the system. For example, when a cognitive system measures an apple, it means that there is a physical interaction between the cognitive system and the apple. As a result, the cognitive system will recognize that this is an apple (e.g., the interaction may be via light and the result of it will be activation of retinal cells which eventually, after sufficient interactions, will lead to the recognition of the apple). In the case of cognitive systems, there are measurements of mental states. This kind of measurement means that the cognitive system interacts with a content-bearing cognitive state, like a representation, using interactions between different parts of the system. It is a strictly physical and public process (accessible for everybody with the right tools). As a result of this definition, the starting point of the argument is with measurements that are non-controversial, i.e., measurements that are public. These measurements include two types, measurements of behavioral reports and measurements of neural representations (like the phenomenal judgement-representations). For example, physical interaction of light between an apple and a cognitive system causes, in the end of a long process of interactions, the activation of phenomenal judgement-representation of an apple and possibly even a behavioral report ("I see an apple").

Let us start from the naturalistic assumption that phenomenal consciousness should have some kind of physical explanation. The physicalism we assume includes matter, energy, forces, fields, space, time, and so forth, and might include new elements still undiscovered by physics. Panpsychism, naturalistic dualism, and illusionism all fall under such a broad physicalism. (It might be, for example, that physics hasn't discovered yet that there is a basic private phenomenal element in addition to the observed known elements. If this element exists, it needs to be part of the broad physicalism.) In addition, let's assume that the principle of relativity holds, i.e., all physical laws in force should be the same in different frames of reference, provided these frames of references agree about all the results of their measurements. Since we have accepted that consciousness is physical (in the broad sense), we can obtain a new equivalence principle for consciousness. Let's assume two agents—Alice, a conscious human being—and a zombie in the form of a complex, artificial cognitive system that delusionally claims to have phenomenal consciousness. Let's call this artificially intelligent zombie "Artificial Learning Intelligent Conscious Entity," or ALICE. ALICE is a very sophisticated AI. It has the capacity to receive inputs from the environment, learn, represent, store and retrieve representations, focus on relevant information, and integrate information in such a manner that it can use representations to achieve human-like cognitive capabilities.

ALICE has direct access to perceptual information and to some of its own cognitive contents, a developed concept of self, the capacity to reflect by creating representations of its internal processes and higher-order representations, and can create outputs and behavioral responses (it has a language system and the ability to communicate). In fact, it was created to emulate Alice. It has the same representations, memories, and dynamics as Alice's cognitive system and as a result, it has the same behavioral responses as Alice. But ALICE is a zombie (we assume) and doesn't have phenomenal consciousness. On the other hand, conscious Alice will agree with all of zombie ALICE's phenomenal judgments. After sufficient time to practice, ALICE will be able to produce phenomenal judgments and reports nearly identical to those of Alice. After enough representational manipulation, ALICE can say, e.g., "I see a fresh little Madeleine, it looks good and now I want to eat it because it makes me happy." It can also reflect about the experience it just had and might say something like, "I just had a tasty Madeleine cake. It reminded me of my childhood, like Proust. There's nothing more I can add to describe the taste, it's ineffable."

Alice and ALICE will agree about all measurements and observations they can perform, whether it's a measurement of their behavior and verbalizations, or even an "inner" observation about their own judgments of their experience, thoughts and feelings. They will not find any measurement that differs between them, although Alice is conscious, and ALICE is not. For example, they can use a Boolean operation (with yes/no output) to compare their phenomenal judgements (e.g., do both agree that they see a Madeleine and that it's tasty?). Now, let's follow in Einstein's footsteps concerning the equivalence principle between a uniformly accelerated system and a system under a uniform gravitational field. Alice and ALICE are two different observers, and the fact that they obtain the same measurements agrees with the conditions of the relativity principle. Accordingly, because both of them completely agree upon all measurements, they are governed by the same physical laws. More precisely, these two observers are cognitive systems, each with its own cognitive frame of reference. For the same input, both frames of reference agree about the outcome of all their measurements, and thus both currently have the same physical laws in force. As a result, these two systems are equivalent to each other in all physical aspects and we can infer physical laws from one system to the other. According to the naturalistic assumption, phenomenal consciousness is part of physics, so the equivalence between the systems applies also to phenomenal structures. Consequently, we can infer that if Alice has phenomenal properties, then ALICE also must have them. Both Alice and ALICE must have phenomenal consciousness! ALICE cannot be a zombie, like we initially assumed, because their systems are physically equivalent. This equivalence makes it impossible for us to speak of the existence of absolute phenomenal properties in the human frame of reference, just as the theory of relativity forbids us to talk of the absolute velocity in a system. For, by knowing that there is phenomenal consciousness in the human frame of reference and by using the broad physicalist premise, we conclude that there are physical laws that enable phenomenal consciousness in the human frame of reference. Because of the equivalence principle, we can infer that the same physical laws will be present also in the supposed "zombie" cognitive system's frame of reference. The conclusion is that if there is phenomenal consciousness in the human frame of reference, then the "zombie" cognitive system's frame of reference must also harbor phenomenal consciousness.

We started from the premise that ALICE is a zombie and concluded that it must have phenomenal consciousness. One of our premises is wrong: either the broad physicalism, the relativistic principle, or the existence of zombies. Most likely the latter is the odd man out, because we can explain these supposed "zombies" using a relativistic, physicalist framework. As a result, although we started from a very broad notion of physicalism and an assumption that the human has phenomenal consciousness and the "zombie" cognitive system does not, the relativity principle forces us to treat phenomenal structures as relativistic. According to the relativity principle, there is no absolute frame of reference; there are only different observers that obtain different measurements. If the observers obtain the same measurements, there cannot be anything else that influences them. There is nothing over and above the observers (no God's-eye-view), and if they observe that they have phenomenal properties, then they have phenomenal consciousness. Because there is nothing over and above the observer, we can generalize this result even further for every cognitive system that has phenomenal judgments: Any two arbitrary cognitive observers that create phenomenal judgment-representations also have phenomenal consciousness. Zombies cannot exist (assuming their cognitive systems create phenomenal judgments). In other words, we obtain an equivalence between functional consciousness (which creates phenomenal judgments) and phenomenal consciousness. Notice that even if we start from the naturalistic dualism of Chalmers and assume a broader physics including phenomenal elements alongside other aspects in the universe, the relativistic principle still forces this kind of broad physics to have the same consequences, viz., that zombies cannot exist and that there is an equivalence between phenomenal judgement-representations and phenomenal properties.

(Formally, phenomenal judgement-representations and phenomenal properties are isomorphic. They have the same mathematical form and there is a transformation between them that preserves this form. See section 2.3. in [the full paper](#)). As a result, the relativistic principle undermines the dualist approach altogether.

**Discussion**

**Consciousness as a Relativistic Phenomenon**

As we saw, the assumption that the relativity principle also includes measurements of cognitive systems forces us to treat consciousness as a relativistic phenomenon. If the relativity principle is true, then zombies are not consistent with the extant laws of nature (note that this is a different claim from logical possibility). Recall that both naturalistic dualism and illusionism can be understood to be opposing responses to the same basic paradox of phenomenal judgment: phenomenal judgments seem to not require phenomenal consciousness. This opens the possibility for zombies. Chalmers (2017; 1996) takes the threat of zombies to force us to accept that phenomenal properties are an additional fundamental component of reality. Illusionists instead take *us* to be the zombies (Frankish, 2017; Graziano et al., 2020). But we have demonstrated that zombies are not possible in a relativistic universe. The illusionist can, of course, point out that we have assumed that Alice is phenomenally conscious, which they do not grant. While this is an assumption (and not a controversial one for any but the illusionist), what we point out is that the illusionist contention that there are zombies, and that *we* are the zombies, is premised on the paradox of phenomenal judgment. But we have eliminated that paradox by eliminating the possibility of zombies in a relativistic universe. Therefore, we have eliminated the illusionist motivation to deny that Alice is phenomenally conscious. While we can't prove that Alice is indeed phenomenally conscious, the illusionist no longer has a reason (viz., the paradox of phenomenal judgment) to doubt it.

As a result of the relativity principle, there is a formal equivalence between functional consciousness (creating phenomenal judgements) and phenomenal consciousness (qualia and eidetic structures; see also the mathematical argument in [the full paper](#)). Ultimately, we are returning to Nagel's (1974) definition of consciousness: that the creature is conscious if there is something that is like to be this creature. If an observer has phenomenal judgements, then the observer measures that there is something that it is like to be that observer, and this observer is conscious. The essence of the relativistic principle is that there is nothing over and above the observer. Hence, in contrast to illusionism and naturalistic dualism, there cannot be any third-person or God's-eye perspective telling us that some observer is delusional and doesn't *really* have consciousness.

Moreover, phenomenal properties are no longer absolute determinations, but depend on the observer's cognitive frame of reference, just as in the case of constant velocity and the question of who's stationary and who's moving. In a non-relativistic universe, an object $O$ evinces either property $P$ or $\neg P$. Properties are absolute determinations. But in a relativistic universe, an object $O$ evinces either property $P$ or $\neg P$ *with respect to some observer $X$*. In the first-person cognitive frame of reference, quale $Q$ (for example) is observable, while any third-person cognitive frame of reference observes $\neg Q$, i.e., that there is no quale.

Cognitive frames of reference are defined by the dynamics of the cognitive systems involved (according to the three-tier information processing model, see the full paper). The first-person frame of reference takes into account also the position of the dynamics (eq. 49–50 in the full paper). The first-person frame of reference of Alice, for example, is the position from within Alice's cognitive system, which is a physical position in space and time where the dynamics of the system take place. This is the only situation that satisfies the condition for the first-person frame of reference (eq. 42 in the full paper). For that reason, the transformation function (that transforms between cognitive frames of reference and between first and third frames of reference) takes into account not only whether the input is recognizable by the cognitive system as one of its own representations, but also if the input is in the position of the cognitive system (eq. 46 in the full paper). The third person frames of reference are the positions of other cognitive systems, like Bob, that measure Alice's cognitive system. While Alice may observe herself feeling happiness as she's looking at a rose, Bob will only measure patterns of neural activity. Recall the case of constant velocity, wherein Alice claims to be at rest while Bob is moving, while Bob claims that Alice is the one moving and that he is stationary. From Alice's perspective, she has qualia and Bob only has patterns of neural activation, while from Bob's perspective Alice just has patterns of neural activation while he has qualia. In other words, just as in the case of constant velocity, Bob and Alice both measure that they are the stationary ones, and hence will not agree on who is stationary and who is moving, there is likewise a relativistic equivalence between all cognitive systems that have phenomenal consciousness or functional consciousness. Both Alice and Bob will measure that they are the ones who have phenomenal consciousness while the other has only neural patterns and hence will not agree on who has phenomenal consciousness.

Alice and Bob will continue to argue over who is right. As a result, Alice might be an illusionist and claim not only that Bob is delusional about having qualia, but moreover that qualia don't exist at all. And, as a response, Bob might claim that qualia are uniquely private and nonphysical phenomena. But, just as with the constant velocity case, their conclusions are wrong because they don't grasp the relativistic principle. The illusionist mistake is to claim that the third-person perspective is the only legitimate perspective. When Alice observes Bob's neural firing, he supposedly should infer that all his "qualia" really are just neural firing. Yet the first-person perspective is also a physical arrangement of a cognitive system. It is the cognitive system from its own observer perspective. Both frames of reference are equivalent, there is no observer position that is privileged, and it is impossible to privilege one observer cognitive position over the other. Alice's perspective is not some absolute, "correct" perspective that sets the standard for measurement. Bob doesn't have phenomenal consciousness relative to Alice's perspective but *does* have phenomenal consciousness relative to Bob's perspective. In other words, Bob evinces ¬$Q$ with respect to Alice, and Bob evinces $Q$ with respect to Bob. This is not a logical contradiction, but a statement of the relativity of properties.

On the other hand, qualia and eidetic structures are not private and hence phenomenal consciousness is neither some mysterious force beyond the realm of science nor an irreducible element of reality. Rather, they *appear* to be private because in order to measure them, one needs to be in the appropriate frame of reference, viz., that of the cognitive system in question (see equations 42, 45 in the full paper). It is ultimately a question of causal power. Only from this frame of reference is there causal power for the representations in the dynamics of the system. Only from the frame of reference of the cognitive system are these neural patterns recognized as representations (eq. 45 in [the full paper](the full paper)). These representations are the input and output of the cognitive system (eq. 2–10 in [the full paper](the full paper)); they are the ones that cause the dynamics in this cognitive frame of reference. From outside that observer reference frame, as in the position the neuroscientist takes as a third-person observer, the same exact phenomena appear as neural computations. This is because the third-person perspective is constitutively outside of the dynamics of representations of the cognitive system in question, and hence these representations do not have any causal power over the neuroscientist (equations 36–38 in [the full paper](the full paper)). According to the equivalence principle, when Alice observes herself to be happy, it is because her cognitive system can recognize and use the appropriate phenomenal judgements, and these judgements are equivalent to phenomenal consciousness (equations 35 in [the full paper](the full paper)), because there is always an equivalent phenomenal judgement representation for every phenomenal property, and we cannot distinguish between them. These representations and their relations cause the cognitive dynamical system to react with new representations and instantiate new relationships between them. As a result, we get a dynamical system that uses *very specific* representations as variables and as outputs. Any other cognitive system, like that of Bob, uses *different* representations and thus cannot use the representations of Alice directly in its own dynamics (equations 36–38 in [the full paper](the full paper)). The only possibility left for Bob is to process Alice's representations through his sensory system and build his own representations. Consequently, we get a sharp difference between the self-measurements of the cognitive system (first-person perspective) and measurements from the outside the cognitive system (third-person perspective), which are mediated solely through the three-tier hierarchy—from the sensation subsystem to the perception subsystem and on to the functional consciousness subsystem (equations 39–40 in [the full paper](the full paper)). Ultimately, the reason for the sharp difference between first-person and third-person frames of reference is because the sensation subsystem can observe only the physical properties of inputs from the outside of the cognitive system, while the functional consciousness subsystem can only observe representations from within the cognitive system (equations 41–43 in [the full paper](the full paper)). Even if the two cognitive systems at hand are in the same cognitive frame of reference (which mean that they have the same dynamics), because they have different spatial positions where their dynamics take place, they will not be in the same first-person frame of reference. They will still need to use their sensation subsystem to measure the other system (hence satisfying the condition of the third-person frame of reference, eq. 43 in [the full paper](the full paper)).

Phenomenal properties are not truly private. They seem private because it is nontrivial to do the transformation to the appropriate cognitive frame of reference. Nevertheless, this kind of transformation is possible in principle. We showed that, as a result of the relativity principle, there is a transformation between the qualia of all cognitive frames of reference (equations 44, 46,51 in [the full paper](the full paper)). Using this transformation, we obtained an equation that agrees with the relativistic principle (equation 44 in [the full paper](the full paper)). It describes the laws of physics with the same form in all admissible cognitive frames of reference. In other words, this form will stay the same regardless of which cognitive frame we choose (it doesn't matter for the equation what the specific $\mu$ and $v$ cognitive frames are). The equation enables us to move from the phenomenal consciousness of one cognitive frame of reference to the phenomenal consciousness of another frame and from first-person frame of reference to third person frame of reference (and vice versa). According to the transformation, each observer can change cognitive frames of reference to any other frame by changing the dynamics of its cognitive system. The transformation equation is built in such a way that ensures that after we applied the transformation and moved from one frame to the other, we will get the correct values of the measurements in the new cognitive frame of reference. For example, the equation ensures that the relations between first- and third-person frames of reference are always satisfied. The observer will always measure qualia and eidetic structures from within its own frame, and brain patterns (or any other physical patterns that govern the cognitive system) for all other frames. The only way to enter the frame of reference of the cognitive system is to have the dynamics of representations of that system, because they only have the right kind of causal power from within that cognitive system (they are the right "fuel" of the cognitive system—the inputs and outputs of the system). Hence, third-person studies will get different measurements than those taken from within the system itself, unless a proper transformation were to change the frame of reference of the third-person cognitive system to have the exact dynamics of the cognitive system in question. After such a transformation, there would be an equivalence between the two systems and they would observe the same conscious experience. Obviously, no such transformation is currently technologically feasible, but, in principle, it is possible. Because of this transformation that preserves the form of the measurements, phenomenal structures (first-person perspective) and phenomenal judgement-representations (third-person perspective) are equivalent and isomorphic to each other. They are not equal, but they have the same preserved mathematical form such that we can map between them. All that separates them are different kinds of measurements (causing different kinds of properties).

**The Hard Problem Dissolves**

The relativistic theory of consciousness dissolves the hard problem. If there is no irreducibly private property of phenomenal features, there is no need for adding new, exotic elements to reality, nor is there a need to explain phenomenal features away as illusory. There is also no explanatory gap, because there is no need to explain how physical patterns in the brain create private, irreducible properties. Because consciousness is a relativistic phenomenon, physical patterns (e.g., of neural representations) and phenomenal properties (e.g., qualia) are two sides of the same coin. Both are valid physical descriptions of the same phenomenon from different frames of reference. Although phenomenal properties supervene on physical patterns (the representations that we called 'phenomenal judgements'), they are not created by the physical patterns. Instead, phenomenal properties are the result of a special measurement of these physical representations by the observer. For this observer, these representations are the ones that cause its own dynamics. Notice that according to the relativistic theory of consciousness, the opposite is also true and physical patterns (phenomenal judgements) supervene on phenomenal properties. They are not created by phenomenal properties; instead, physical representations are the result of a measurement of these phenomenal properties by a different observer. For this cognitive frame of reference, these phenomenal properties have no causal power over its dynamics. Phenomenal and physical aspects supervene on each other. There is a subtle identity between them (i.e., they are equivalent). They are just different perspectives of the same phenomenon from different cognitive frames of reference.

As a result, the relativistic theory of consciousness is a physicalist theory, but not a reductive theory, because there is no reduction of phenomenal properties to brain patterns. For phenomenal properties we need a cognitive system with the right kind of representations and the right kind of measurement.

In the introduction, we raised a concern about whether any physicalist theory can solve the hard problem of consciousness. How can there be an identity between public properties like structure and dynamics and between private qualia? The answer is that we didn't include relativity as a part of physicalism. The relativistic theory of consciousness shows that phenomenal states are not private and gives an explanation of why they are different from other physical states. Different kinds of measurements give rise to different properties. While our sensation subsystem can only directly measure a physical substrate via measurement devices like the retina, the perceptual and cognitive subsystems directly measure only the roles and relations of representations within their subsystems and cannot directly measure the physical substrate serving as the referent of their representations (that's why in the equations that describe the dynamics of the perceptual and cognitive subsystems, the inputs are representations and not the physical substrate of the representations, while in the sensation system the inputs are the physical substrates themselves like light or sound waves. See [the full paper](#)). From these different kinds of measurements different kinds of properties arise.

Chalmers (1996; 2017) parses the zombie argument in terms of the logical possibility of zombies. He would not deny that zombies are inconsistent with the extant laws of nature, because he speaks only about logical possibility in general. However, the relativistic theory of consciousness carries ontological constraints and conditions about the existence of consciousness in every possible world. For example, the difference between first- and third-person frames of reference arise because of the difference between the kind of measurements that can be taken within each frame of reference. There could be a possible world, not governed by the known laws of nature, wherein the relativistic principle is false. But then, there would be no different frames of reference, and the different kinds of measurements would yield the same result. In other words, there would be no first-person and third-person frames of reference. In such a possible world, we would be left with either the illusionist view, with no consciousness at all (i.e., everybody is a zombie with just third-person perspective), or we would be left with a Berkeleian idealist world where there are only phenomenal properties (with shared first-person perspective for everything). Consequently, according to our relativistic theory of consciousness, there cannot be a world that has both phenomenal consciousness *and* zombies. This conclusion undermines the zombie argument. The main goal of the zombie argument was to establish that phenomenal properties cannot be explained reductively in terms of physics. But the relativistic theory does just that, demonstrating that phenomenal consciousness and physics can be reconciled. There might be a logical possibility of zombies, but in such possible worlds, the physical mechanism that allows the transformation to phenomenal properties (i.e., the relativistic principle) must be absent and hence consciousness must be absent all together as well (because all there is left in such a world is the physical substance that constitutes the world). In sum, in non-relativistic worlds without different frames of reference, we could have *either* illusionism (only physical properties) *or* Berkeleian idealism (only phenomenal properties). But in a relativistic world where different frames of reference measure different properties, *both* physical *and* phenomenal properties are possible for the same entity.

Phenomenal consciousness is not private—all we need to do to measure phenomenal properties is to change our perspective by moving to the appropriate frame of reference, i.e., the frame wherein the representations have causal power (i.e., the frame that measures the representations directly). As a result, this frame doesn't measure them merely as physical patterns, but as what they represent and stand for. For example, the representation of an apple has some causal power in the system according to what the system has learned about apples. The representation will trigger, for example, memories, emotions, and motor processes. For the cognitive system, this is what it means to be an apple. Thus, the physical properties of the representation are not being directly measured, but rather only its role and relations with other representations in the system. Consequently, this representation will be measured by the system as an apple and not as a representation of an apple. That is to say, because there is nothing above and beyond the observer, and because the cognitive system (the observer) measures this representation according to what it does (the representation describes all the functions, properties and relationships of an apple in the system), this representation will be measured by the system as an apple. (That's why phenomenal consciousness is characterized by transparency, i.e., we don't perceive representations but directly perceive things.) This is a direct measure of the representation itself. In the case of phenomenal judgements, these new properties are being measured as phenomenal properties. Specifically, when there is a cognitive frame of reference that creates functional consciousness (phenomenal judgements representations), it is sensible to assume that such a system will have a special subsystem specializing in such complex representations. It uses the complex representations of functional consciousness for its dynamics "(i.e., phenomenal judgment representation, see eq. 5 in the full paper). As a result, these representations have causal power in this frame of reference. The relativity principle ensures us that because no measurement can distinguish between phenomenal properties and their corresponding phenomenal judgements, this cognitive frame will measure these phenomenal judgements as qualia and eidetic structures (equation 35 in the full paper). In sum, according to the relativistic theory of consciousness, for phenomenal structures we need a cognitive system with the right kind of representations and the right kind of measurement. Phenomenal structures are the direct measurements of the complex representations that we called phenomenal judgments.

The relativistic theory of consciousness suggests a solution for the hard problem based in relativistic physics. There are still several open questions that need to be addressed in the future. For example, what are the necessary and sufficient conditions for a cognitive frame of reference to have phenomenal consciousness? And what empirical predictions does this theory yield?
The implications of such necessary and sufficient conditions are huge. According to these condiotions the theory would be able to determine which animal was the first animal in the evolutionary process to have consciousness, when a fetus or baby begins to be conscious, which patients with consciousness disorders are conscious, and which AI systems already today have a low degree (if any) of consciousness.